\documentclass[a4paper]{jpconf}
\usepackage{graphicx}
\begin{document}
\title{Two-junction superconductor-normal metal single-electron trap in a combined on-chip RC environment}

\author{Sergey V. Lotkhov and Alexander B. Zorin}

\address{Physikalisch-Technische Bundesanstalt, 38116 Braunschweig, Germany}

\ead{sergey.lotkhov@ptb.de}

\begin{abstract}
Dissipative properties of the electromagnetic environment as well as
on-chip RC filtering are shown to suppress random state switchings in the
two-junction superconductor(S) - normal metal(N) electron trap. In our
experiments, a local high-ohmic resistor increased the hold time of the
trap by up to two orders of magnitude. A strong effect of on-chip noise
filtering was observed for different on-chip geometries. The obtained
results are promising for realization of the current standard on the basis
of the S-N hybrid turnstile.
\end{abstract}

\section{Introduction}
Experimental realization and reliable control of macroscopic quantum
states require an efficient decoupling of the quantum circuit from
external fluctuations. At low enough temperatures, typically $T \le
100$~mK, the role of equilibrium thermal fluctuations vanish and the
dominant fluctuation mechanisms originate from residual noise photons
approaching the tunneling circuit from its electromagnetic environment
\cite{AverinLikharev91,IngoldNazarov}. $Photon$-$assisted$ $tunneling$ was
regarded as a noise mechanism limiting the metrological accuracy of a
normal-state single-electron tunneling (SET) pump \cite{KautzPAT00}. In
the superconducting state, $quasiparticle$ $excitations$ were found to
provide an influent source of decoherence, for example, in the
Josephson-junction qubits \cite{LangMartinis}. More recently,
$environmentally$ $assisted$ $tunneling$ (EAT) was reported to be
responsible for the accuracy of a hybrid superconductor(S) - normal
metal(N) single-electron turnstile \cite{PekolaEA10}.

On-chip filtering elements, namely a capacitively coupled ground plane and
a local high-ohmic resistor, were recently demonstrated to enhance the
subgap current suppression in single SIN junctions, where "i" stands for
"insulator",  and in a two-junction hybrid SINIS turnstile
\cite{PekolaEA10,LotkhovRturn09}. In our previous work
\cite{LotkhovRSINISTrap11}, a Cr resistor was found to expel the
quasiparticle leak in the turnstile down to the rare single-electron
escape events monitored in real time by an SET electrometer. Our present
study addresses both partial contributions and the integral effect of the
aforementioned improvements in a combined RC environment of a two-junction
SINIS trap. Different sample geometries are analyzed with regard to the
suppression of noise-activated tunneling and possible propagation ways of
the noise photons.

\section{Experiment and results}

\begin{figure}[t]
\begin{minipage}{20pc}
\includegraphics[width=3.4in]{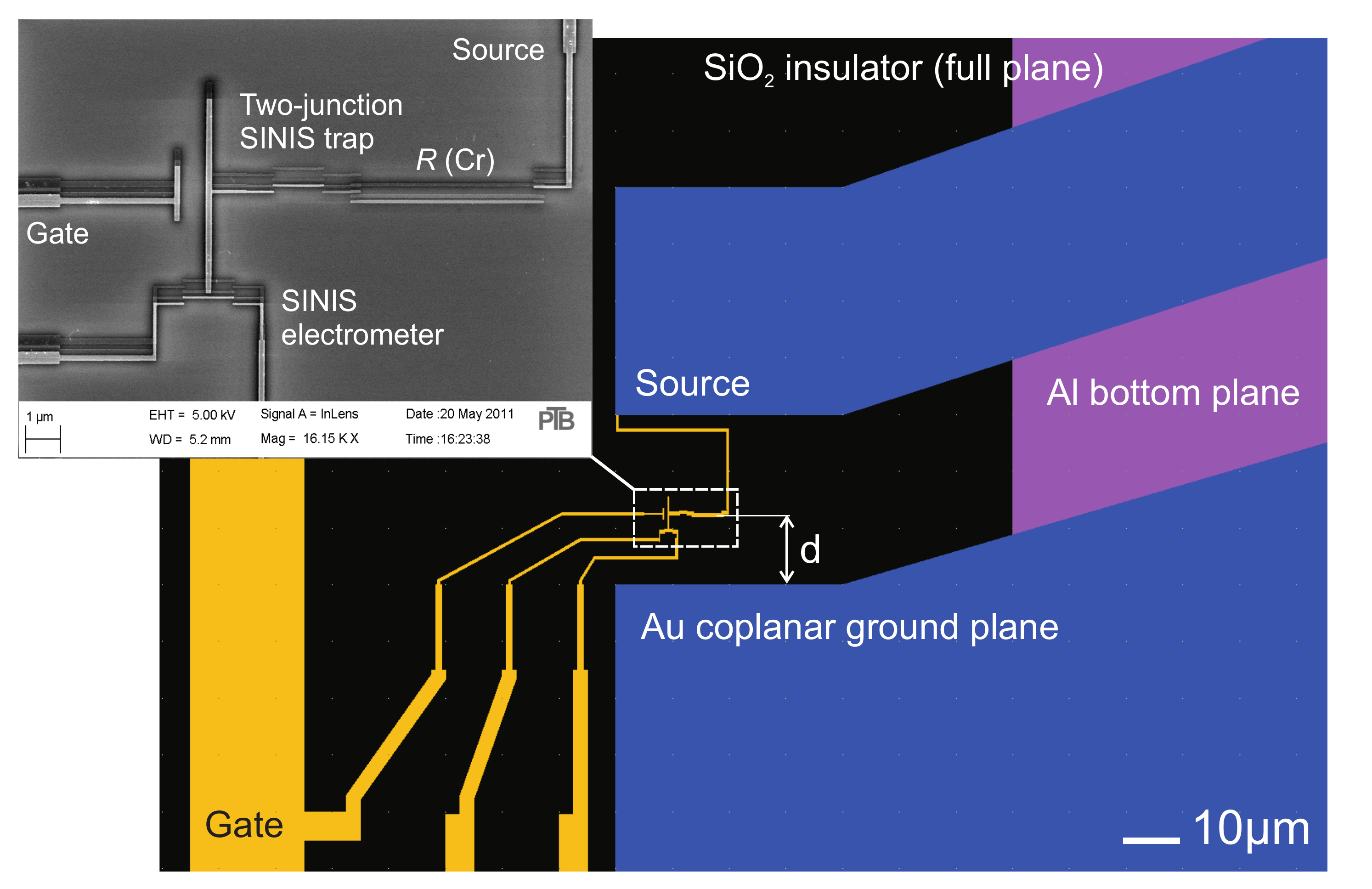}
\caption{On-chip geometry of Sample~5. The inset shows an SEM image of the trapping device
made with SIN junctions of type Al/AlO$_{\rm x}$/AuPd.}
\label{fig1}
\end{minipage}\hspace{2pc}%
\begin{minipage}{16pc}
\includegraphics[width=2.6in]{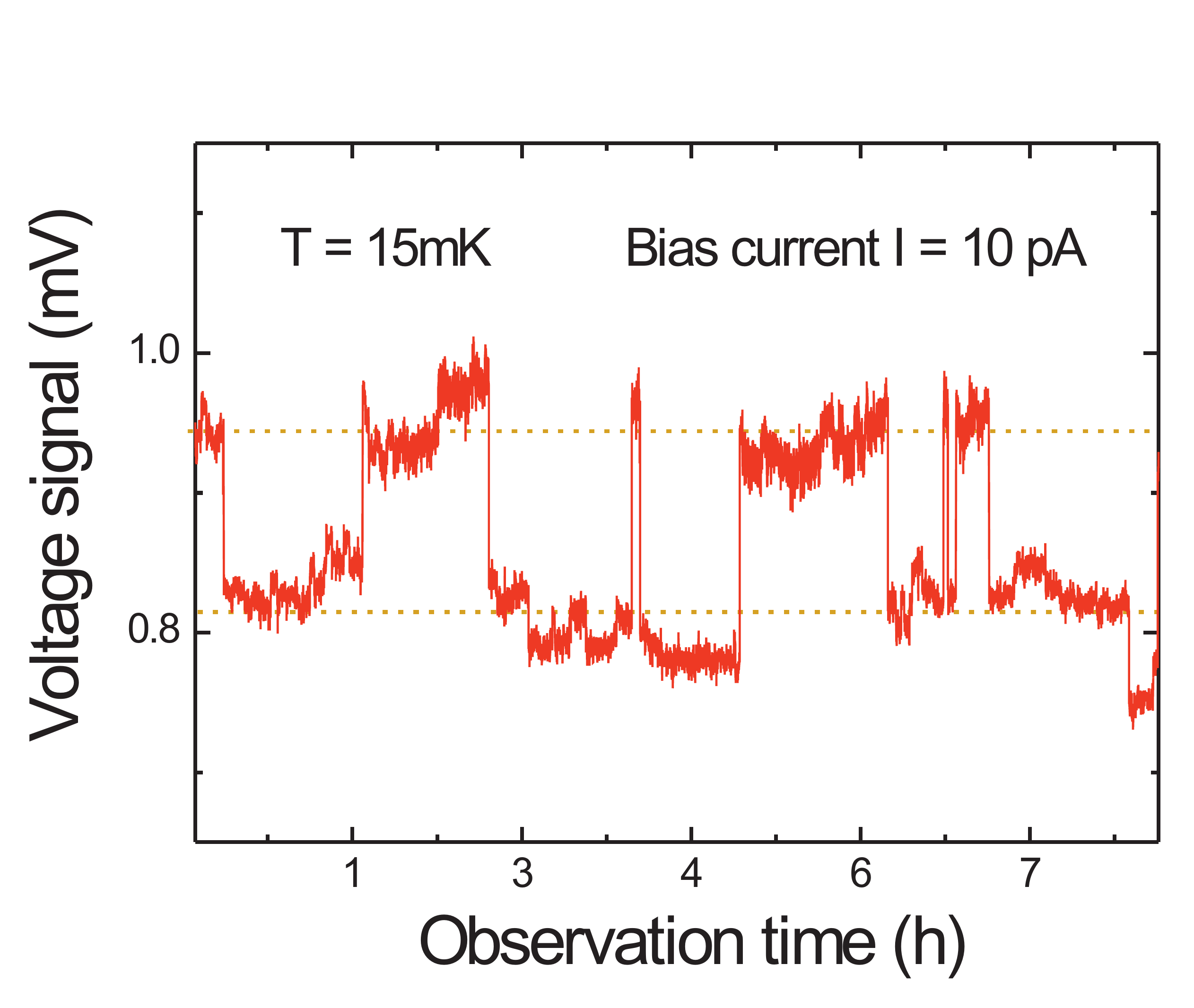}
\caption{A switching track recorded for Sample~5 in the maximum of
the Coulomb blockade in the trap. The dashed lines are eye-guides for the
two neighboring states.} \label{fig2}
\end{minipage}
\end{figure}

Our trapping device, shown in the inset to Fig.~\ref{fig1}, included a
two-junction SINIS turnstile, either with or without a high-ohmic Cr
resistor on the source side, and a dc-SET SINIS electrometer. The
electrometer was coupled capacitively to the opposite, open-end terminal
of the turnstile, used as an electron trapping reservoir with discrete
states. All four SIN(NIS) junctions were nominally of the same area, which
was used in estimating the parameters of the turnstile: the charging
energy $E_{\rm C} \approx e^2/\left ( 2 \times 2 \times C_{\rm T} \right
)$, where $C_{\rm T}$ is the capacitance of a tunnel junction, the tunnel
resistance $R_{\rm T}$, and the superconducting gap of the Aluminum S
leads $\Delta \approx 250~\mu$eV, based on those data obtained for the
electrometer. Similar to the experiment in
Ref.~\cite{LotkhovRSINISTrap11}, the electrometer was used to record the
random state switchings of the trap, see an example in Fig.~\ref{fig2},
and, in this way, to quantify the electron escape process across the
Coulomb energy barrier in the turnstile.  In particular, the average
switching interval (hold time) $\tau_{\rm {max}}$ was assumed to be
inversely proportional to the escape rate and, determined for the trap
adjusted to its maximum Coulomb barrier $\Delta E \sim E_{\rm C}$, could
be compared with our follow-up of the EAT model developed in
Refs.~\cite{IngoldNazarov,PekolaEA10}.

The devices under test were fabricated on thermally oxidized silicon
substrates using the three-shadow evaporation process described previously
in Ref.~\cite{LotkhovRSINISTrap11} and the references therein. That
formerly developed process included the fabrication of a 50~nm-thick
coplanar ground plane (CGP) of Au, which was also common for all the
samples reviewed here (see Fig.~\ref{fig1}). For the present study, every
circuit - except the Sample~4 - was built on top of a 50~nm-thick bottom
plane (BP), either of Al or AuPd, covered by a 200~nm-thick insulating
layer of SiO$_2$. The capacitance $C_{\rm L} \sim $100~pF of each dc lead
to the bottom plane together with the lead resistance $R_{\rm L} \sim
100~\Omega$, formed an effective low-pass filter for the external noise
with the cutoff frequency, $f \sim $~10---100~MHz, well below the
excitation threshold for quasiparticles $f_{\rm {qp}} = \Delta/h
\sim$~50~GHz. The trapping devices were positioned above an opening in the
bottom plane at different distances $d$ to the closest grounding element,
as shown in Fig.~\ref{fig1} and in Table~\ref{table1}. The resistance $R$
of our 5~$\mu$m-long Cr resistor was estimated using an identical test
resistor on the same chip. For the circuits without a Cr resistor, the
lead impedance was accepted being on the scale of 1~k$\Omega$.

\begin{table}[t]
\caption{\label{table1} Parameters of the samples: hold time $\tau_{\rm {max}}$ vs. $E_{\rm C}, R_{\rm T}$ and $R$.}
\begin{indented}
\lineup
\item[]\begin{tabular}{@{}lllllll}
\br
Sample No. & $E_{\rm C},~\mu eV$ & $R_{\rm T},~k\Omega$ & $R,~k\Omega$ & CGP/BP & $d, \mu$m & $\tau_{\rm {max}},~s$\\
\mr

  1a  & \0250 & \0150 & 150     & CGP+BP(Al)    & 200 & \m0.3 \\
  1b  & \0250 & \0150 & $\sim$1 & CGP+BP(Al)    & 200 & $<$0.05$^{\rm a}$ \\
\\
  2a  & \0250 & \0400 & 150     & CGP+BP(AuPd)  & 200 &\0 1.0 \\
  2b  & \0250 & \0400 & $\sim$1 & CGP+BP(AuPd)  & 200 &\0 0.2 \\
\\

  3a  & \0500 & \0800 & 150     & CGP+BP(AuPd)  & 200 & \m500 \\
  3b  & \0500 & \0800 & $\sim$1 & CGP+BP(AuPd)  & 200 & \m2--5 \\
\\
  4   & 1000  & 1850   & 440     & CGP           & 200  & \0200 \\
  5   & \0300 & \0400  & 550     & CGP+BP(Al)    & \010 & 2200 \\

\br
\end{tabular}
\item[] $^{\rm a}$The frequency of switchings exceeded the detection bandwidth of the dc electrometer.
\end{indented}
\end{table}

The experimental results are summarized in Table~\ref{table1}. They were
obtained at the base temperature of our dilution fridge $T =$~15~mK, using
a microwave-tight (but not hermetically-sealed) sample holder equipped
with 1---1.5~m-long Thermocoax$\texttrademark$ coaxial filters per each
dc-line. The structures indexed as "a" and "b", located on the same sample
and differing only by the value of $R$, were measured within the same
low-temperature cycle, which made possible a direct comparison of an
effect of the low- and high-ohmic environment.

\section{Model and discussion}

\begin{figure}[t]
\includegraphics[width=3.0in]{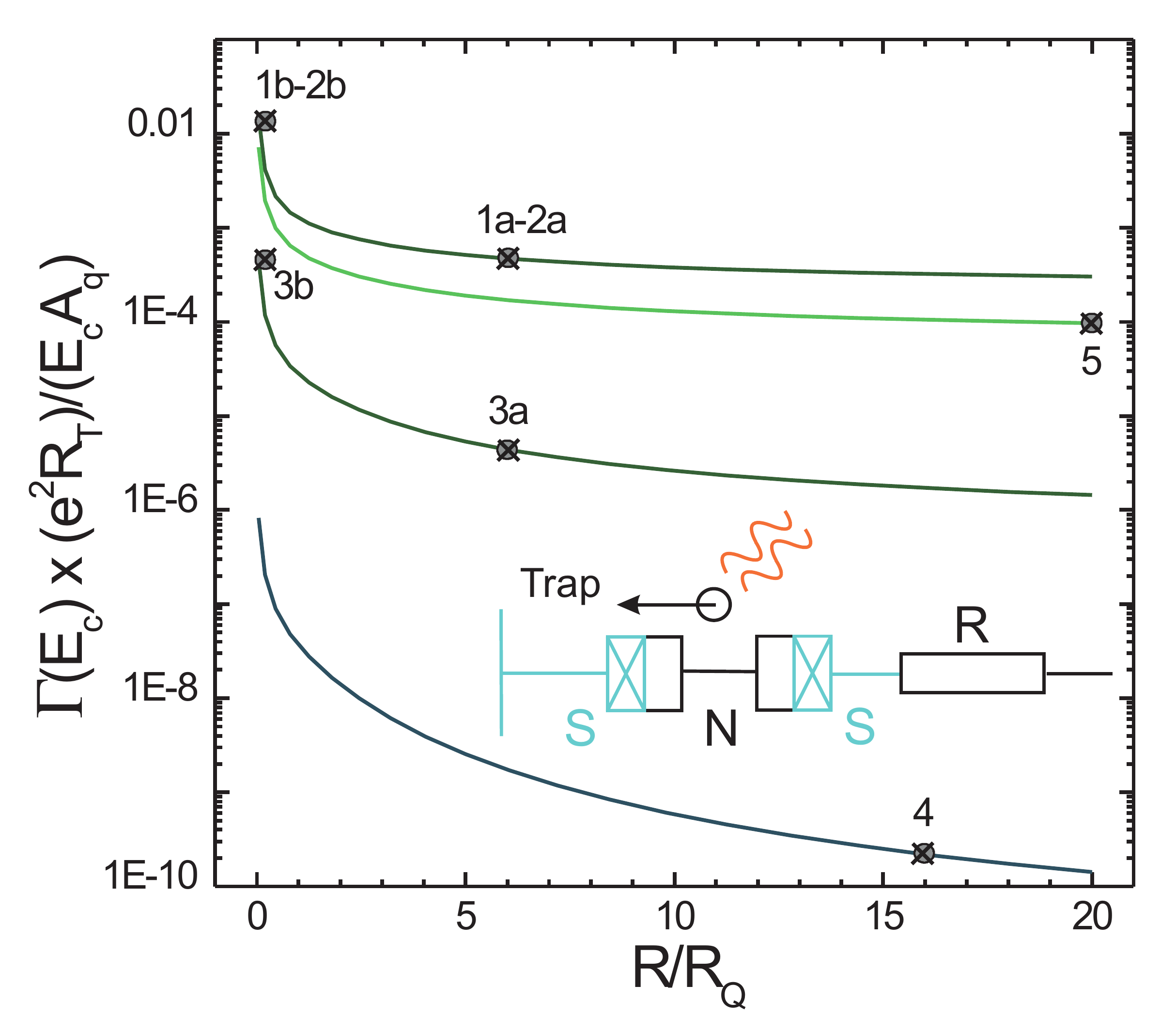}\hspace{2pc}%
\begin{minipage}[b]{18pc}
\caption{Normalized tunneling rate as a function of the ratio
$R/R_{\rm Q}$, where $R_{\rm Q} \equiv h/e^2 \approx$~25.8~k$\Omega$
is the resistance quantum, calculated for $E_{\rm
C}$~=~250,~300,~500, and 1000~$\mu$eV from top to bottom and
$T^{*}$=~1~K. The crossed circles indicate the parameters of the
samples under study.} \label{fig3}
\end{minipage}
\end{figure}

We interpret our data in the framework of the EAT model
\cite{PekolaEA10,LotkhovRSINISTrap11}. We assume a black-body radiation
noise of the environment, equilibrated at the temperature level of the
closest radiation shield, $T^{*}$~$\sim$~1~K, being the most probable
source of higher-energy excitations. It is further assumed that the S
leads are originally free of quasiparticles (see, e.g.,
Ref.~\cite{SairaAluminium}), and the dominant recharging cycles start with
an electron tunneling from the N island, as depicted in the inset in
Fig.~\ref{fig3}, and producing a quasiparticle in the S node of the trap.
This energetically unfavorable process is stimulated by the voltage noise
across the SIN junctions, and we model the noise effect by introducing a
phenomenological non-equilibrium population of states $E$ above the
Fermi-level in the normal metal, $F_{\rm ph}(E,T^{*}) = A_{\rm q} \times
exp(-E/k_{\rm B}T^{*})$. In our approximate model, the dimensionless noise
intensity pre-factor $A_{\rm q}$ is accepted being independent on the
parameters of the trapping device itself, but solely related to the noise
properties of the environment.

For the rate calculation, we make use of the very low temperatures of both
S and N leads, $T_{\rm S}, T_{\rm N} \ll T^{*},\Delta/k_{\rm B}$, which
reduces the golden-rule tunneling formula to the simplified
zero-temperature case:

\begin{equation}
\fl \qquad \Gamma(E_{\rm C}) = \frac{1}{e^2R_T}
\int_{\Delta + E_{\rm C}}^\infty dE_{\rm n}
F_{\rm ph}(E_{\rm n},T^{*})
\int_{\Delta}^{E_{\rm n}-E_{\rm C}} dE_{\rm s}
n_{\rm s}(E_{\rm s}) P(E_{\rm n}-E_{\rm s}-E_{\rm C}),\label{eq:goldenrule}
\end{equation}
where $n_{\rm s}(E) = \frac{E}{\sqrt{E^2 - \Delta^2}}$ is the BCS density
of states in a superconductor and the function $P(E)$, the absorption
spectrum of the environment, is that for the pure ohmic environment $R$ at
zero temperature \cite{IngoldNazarov}. An electron escape across the
turnstile is completed by tunneling through the second SIN junction at
much higher rate, thus making the onset rate defined in equation
\ref{eq:goldenrule} the dominant term in the overall escape rate:
$\tau_{\rm {max}}^{-1} \approx \Gamma(E_{\rm C})/2$. The tunneling rate
$\Gamma$ appears in equation~\ref{eq:goldenrule} proportional to the noise
pre-factor $A_{\rm q}$, so the hold time of similar trapping devices
should provide a direct measure of the on-chip noise suppression through
$\tau_{\rm {max}} \propto A_{\rm q}^{-1}$.

Figure~\ref{fig3} predicts a reduction of the tunneling rate by increasing
the resistance $R$. There is a reasonable agreement with the hold times in
Samples~3a and 3b and still quantitative deviation from the data obtained
for Samples~1a,b and 2a,b. In particular, an increase up to $\tau_{\rm
{max}} \sim$~500~s due to $R$~=~150~k$\Omega$ is as high as by two orders
of magnitude between Samples~3a and 3b. Within the sample pairs 1a,b and
2a,b with a lower value of $E_{\rm C} \approx$~250~$\mu$eV and $\tau_{\rm
{max}} \le$~1~s, the model predicts factor 3 smaller ratios of the hold
times. Nevertheless, the experimental ratios were found to be even smaller
than the predicted ones, presumably, due to the finite frequency bandwidth
of our detector.

A pronounced effect of the bottom plane was observed in the comparison of
Sample~5 to Sample~4. Taking into account the calculated data shown in
Fig.~\ref{fig3} as well as the experimental data in Table~\ref{table1},
one can conclude on the ratio $A_{\rm q}^{\rm {CGP}}/A_{\rm q}^{\rm
{CGP+BP(10~\mu m)}}$$\sim$~$10^7$ as an effect of the bottom plane. A
further comparison of Samples~5 and 3a, positioned at a larger distance
$d$ to the large-area planes, indicates noise suppression by factor
$A_{\rm q}^{\rm {CGP+BP(200~\mu m)}}/A_{\rm q}^{\rm {CGP+BP(10~\mu
m)}}$$\sim$~$10^2$ in vicinity of the shielding planes. Since for both
samples the on-chip filtering design is identical, we conclude on the
space propagation channel for the noise photons. Finally, we note that the
derived relative values of $A_{\rm q}$ depend exponentially on the
measured uncertainties in $E_{\rm C}$ and $T^{*}$ and need more direct
confirmation.

To conclude, a hold time of an SINIS trap was measured for various
specially engineered on-chip electromagnetic environments. The combined
effect of both the high-ohmic resistor and the improvement in the sample
geometry (Samples 2b and 5) manifested itself in the hold time increase by
about four orders of magnitude. Strong partial effects of both the
high-ohmic resistor and the on-chip RC filtering (bottom plane) were
observed. Further experiments are in progress to quantify the effect of
the sample design on the noise-activated tunneling processes.

\ack
Fruitful discussions with J~Pekola and A~Kemppinen are gratefully
acknowledged. Technological support from T~Weimann and V~Rogalya
is appreciated. The research conducted within the EU project SCOPE
received funding from the European Community's Seventh Framework
Programme under Grant Agreement No. 218783.


\section*{References}
\begin{thebibliography}{8}

\bibitem{AverinLikharev91} Averin D V and Likharev K K 1991 {\it Mesoscopic Phenomena in Solids}
(Amsterdam: Elsevier) pp~173-271

\bibitem{IngoldNazarov} Ingold G L and Nazarov Yu V 1992 {\it Single Charge Tunneling, Coulomb Blockade Phenomena in Nanostructures}
(NATO ASI Series B vol~294)(New York: Plenum Press) ch 2

\bibitem{KautzPAT00} Kautz R L, Keller M W and Martinis J M 2000 {\it Phys Rev B} {\bf 62} 15888

\bibitem{LangMartinis} Lang K M, Nam S, Aumentado J, Urbina C and Martinis J M 2003 {\it IEEE Trans. Appl. Superconduct.} {\bf 13} 989

\bibitem{PekolaEA10} Pekola J P, Maisi V F, Kafanov S, Chekurov N, Kemppinen A,
Pashkin Yu A, Saira O-P, M\"ott\"onen M and Tsai J S 2010 {\it Phys Rev Lett } {\bf 105} 026803

\bibitem{LotkhovRturn09} Lotkhov S V, Kemppinen A, Kafanov S, Pekola J P and Zorin A B 2009 {\it Appl Phys Lett } {\bf 95} 112507

\bibitem{LotkhovRSINISTrap11} Lotkhov S V, Saira O-P, Pekola J P and Zorin A B 2011 {\it New Journ Phys} {\bf 13} 013040

\bibitem{SairaAluminium} Saira O-P, Kemppinen A, Maisi V F and Pekola J P 2011 Is aluminum a perfect superconductor? {\it arXiv:1106.1326v1}

\end{thebibliography}
\end{document}